\documentclass{article}

\usepackage{arxiv}

\usepackage[utf8]{inputenc} 
\usepackage[T1]{fontenc}    
\usepackage{hyperref}       
\usepackage{url}            
\usepackage{booktabs}       
\usepackage{amsfonts}       
\usepackage{nicefrac}       
\usepackage{microtype}      
\usepackage{lipsum}		
\usepackage{setspace}
\usepackage{graphicx}
\usepackage{doi}
\usepackage{amsmath}
\usepackage[numbers,sort&compress]{natbib}

\title{A Three-Dimensional Dodecaphenylyne-Derived Carbon Allotrope with Anisotropic and Auxetic-Like Mechanical Behavior}

\author{%
\parbox{0.95\linewidth}{\centering
\href{https://orcid.org/0000-0003-4699-5886}{\includegraphics[scale=0.09]{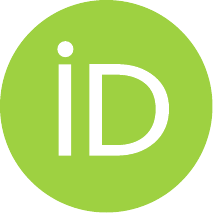}\hspace{1mm}}Kleuton A.~L.~Lima\textsuperscript{1},
\href{https://orcid.org/0000-0002-8366-7227}{\includegraphics[scale=0.09]{icons/orcid.pdf}\hspace{1mm}}Jos\'e A.~dos S.~Laranjeira\textsuperscript{2},
\href{https://orcid.org/0000-0001-5048-0696}{\includegraphics[scale=0.09]{icons/orcid.pdf}\hspace{1mm}}Bill D. A. Huacarpuma\textsuperscript{3},
\href{https://orcid.org/0000-0001-7653-0428}{\includegraphics[scale=0.09]{icons/orcid.pdf}\hspace{1mm}}Nicolas F.~Martins\textsuperscript{2},
\href{https://orcid.org/0000-0002-5217-7145}{\includegraphics[scale=0.09]{icons/orcid.pdf}\hspace{1mm}}Julio R. Sambrano\textsuperscript{2},
\href{https://orcid.org/0000-0003-0145-8358}{\includegraphics[scale=0.09]{icons/orcid.pdf}\hspace{1mm}}Douglas S.~Galv\~ao\textsuperscript{3},
and
\href{https://orcid.org/0000-0001-7468-2946}{\includegraphics[scale=0.09]{icons/orcid.pdf}\hspace{1mm}}Luiz A.~Ribeiro Jr\textsuperscript{2,$\dag$} \\
\vspace{0.6em}
{\normalfont\normalsize
\textsuperscript{1}Department of Applied Physics and Center for Computational Engineering and Sciences, State University of Campinas, Campinas, 13083-859, SP, Brazil\\
\textsuperscript{2}Modeling and Molecular Simulation Group, S\~ao Paulo State University (UNESP), School of Sciences, Bauru 17033-360, SP, Brazil\\
\textsuperscript{3}Computational Materials Laboratory, LCCMat, Institute of Physics, University of Bras\'ilia, 70910-900, Bras\'ilia, Federal District, Brazil\\
\vspace{0.6em}
\href{https://scholar.google.com.br/citations?user=e_2ul00AAAAJ&hl=pt-BR}{\includegraphics[scale=0.05]{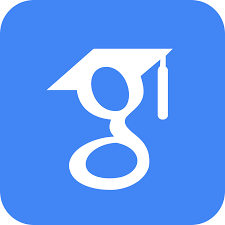}} \href{https://www.linkedin.com/in/kleuton-antunes-1023b4392/}{\includegraphics[scale=0.05]{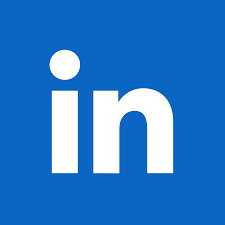}}\hspace{0.1cm}\texttt{\textsuperscript{*}kleuton@unicamp.br} \\
\vspace{0.1cm}
\href{https://scholar.google.com/citations?user=EgsxcaUAAAAJ\&hl=pt-BR}{\includegraphics[scale=0.05]{icons/gscholar.png}} \href{www.linkedin.com/in/luiz-ribeiro-164221225}{\includegraphics[scale=0.05]{icons/linkedin.png}}\hspace{0.1cm}\texttt{\textsuperscript{$\dag$}ribeirojr@unb.br}\\
}
}%
}

\renewcommand{\shorttitle}{Kleuton A.~L.~Lima et al.}

\hypersetup{
pdftitle={3d-dodeca},
pdfsubject={3d-dodeca},
pdfauthor={Kleuton A. L. Lima, Luiz A. Ribeiro Jr},
pdfkeywords={Two-dimensional metals, Kagome lattice, Goldene, Lattice dynamics, Strain engineering, First-principles calculations.},
}

\begin{document}
\maketitle

\onehalfspacing

\begin{abstract}
We introduce 3D-DPhyne, a novel three-dimensional (3D) carbon allotrope derived from the dodecaphenylyne framework, and investigate its structural, electronic, optical, and mechanical properties using first-principles calculations. The proposed structure forms a tetragonal, topologically complex network of four-, six-, and twelve-membered carbon rings with mixed sp/sp$^{2}$ hybridization and a formation energy of $-7.87$~eV/atom, comparable to other stable carbon allotropes. Phonon dispersion calculations show no imaginary modes, and \textit{ab initio} molecular dynamics simulations at 1000~K confirm robust thermal stability without bond breaking. Electronic structure analysis reveals metallic character, with multiple bands crossing the Fermi level and dominant contributions from carbon $p$ orbitals, consistent with a fully delocalized 3D $\pi$-conjugated network. The optical response is anisotropic, exhibiting strong absorption in the visible and ultraviolet regions and low reflectivity across a broad range of photon energies. Mechanical analysis reveals pronounced elastic anisotropy, with Young's modulus varying from approximately 40 to 490~GPa depending on direction. Poisson's ratio displays unconventional directional behavior, including auxetic-like responses.
\end{abstract}


\section{Introduction}

Carbon exhibits unparalleled versatility in forming stable structures with diverse dimensionalities, bonding configurations, and electronic properties \cite{xiao2022dimensionality,speranza2019role}. This versatility has led to the discovery of a wide range of allotropes with remarkable physical characteristics \cite{nasir2018carbon,tiwari2016magical}. Two-dimensional (2D) carbon systems, most notably graphene \cite{geim2009graphene}, have attracted sustained attention for their exceptional mechanical strength, high carrier mobility, and unique electronic properties \cite{tiwari2016magical,yadav2024carbon}. Nevertheless, graphene has intrinsic limitations, such as its zero band gap and limited tunability of electronic and optical responses \cite{rani2013designing}. These have motivated extensive efforts to design alternative carbon architectures with tailored properties \cite{enyashin2011graphene,lusk2009creation,LIMA2025117868,lima2025structural,lima2025petal,lima2025first,mortazavi2023electronic,bafekry2024layered,bafekry2022theoretical,bafekry2022tunable,xu2013graphene}. As a result, researchers have synthesized and predicted various non-honeycomb \cite{fan2021biphenylene,hou2022synthesis,aliev2025planar,desyatkin2022scalable} and porous \cite{zhang2019recent} 2D carbon allotropes. In these materials, structural complexity and topology play central roles in determining functionality.

Beyond two dimensions, three-dimensional (3D) carbon frameworks introduce additional degrees of freedom for tailoring electronic dispersion, enhancing mechanical rigidity, and tuning optical responses \cite{shehzad2016three,liu2020carbon}. Although several 3D carbon phases have been suggested \cite{fayos1999possible,fu2017progress,sun20203d}, many rely on simple bonding patterns or lack the intricate topological features characteristic of novel 2D multiring networks. The transformation of complex 2D carbon lattices into three dimensions remains largely unexplored \cite{felix2025novel,tromer2025electronic,felix20253d,tromer2024transforming,lebedeva2007kinetics,fayos1999possible}, primarily because of the promise of enabling dimensional crossover effects, facilitating 3D $\pi$-electron delocalization, and yielding inherently anisotropic properties. Investigating such 3D architectures offers a route to discover carbon materials with unconventional structure–property relationships.

Recently, dodecaphenylyne has been proposed as a 2D carbon allotrope characterized by a multiring framework composed of four-, six-, and twelve-membered rings, exhibiting pronounced anisotropy and a conjugated electronic structure \cite{lima2026first}. Building on this topology, we numerically introduce a 3D carbon allotrope, 3D-DPhyne, derived from the dodecaphenylyne framework. We show that 3D-DPhyne combines dynamic and thermal stability with metallic electronic behavior and pronounced anisotropy in both optical and elastic responses. 

In this work, we perform a comprehensive first-principles investigation of the structural and physical properties of 3D-DPhyne using density functional theory (DFT) and ab initio molecular dynamics (AIMD) simulations. We analyze its crystallographic symmetry, energetic stability, vibrational spectrum, and thermal robustness, establishing the viability of the proposed 3D framework. We further examine the electronic band structure, orbital characteristics, optical response, and elastic properties to elucidate the impact of dimensional extension on charge delocalization, metallic behavior, and pronounced anisotropy induced by the lattice topology. By systematically correlating structural features with electronic, optical, and mechanical responses, this study provides insights into the role of dimensional crossover in complex carbon networks, such as dodecaphenylyne. 

\section{Methodology}

All calculations were performed using DFT with the CASTEP package \cite{clark2005first}. We used norm-conserving pseudopotentials and the Perdew–Burke–Ernzerhof (PBE) exchange–correlation functional in the generalized gradient approximation (GGA) \cite{perdew1996generalized}. Atomic positions and lattice parameters were optimized with the Broyden–Fletcher–Goldfarb–Shanno (BFGS) algorithm \cite{head1985broyden,PFROMMER1997233}. We set convergence criteria at $1.0\times10^{-5}$~eV for total energy, $1.0\times10^{-3}$~eV/\r{A} for maximum residual force, and $1.0\times10^{-2}$~GPa for maximum residual stress.

Following structural optimization, Brillouin-zone integrations were performed using a Monkhorst–Pack $5\times5\times1$ $k$-point mesh during relaxation, while a denser $20\times20\times1$ grid was adopted for electronic structure and optical property calculations. To avoid spurious interactions between periodic images, a vacuum spacing of 20~\r{A} was introduced along the non-periodic direction.

Building on electronic structure calculations, the vibrational properties were investigated using density functional perturbation theory (DFPT). Phonon dispersion relations were computed on a $5\times5\times1$ $q$-point mesh, employing a finite displacement step of 0.05~\AA$^{-1}$ and a force convergence threshold of $1.0\times10^{-5}$~eV/\AA$^{2}$. Thermal stability was further assessed using AIMD simulations at 1000~K in the canonical (NVT) ensemble.

To further characterize the material, elastic properties were obtained by applying small uniaxial strains to the optimized structure and calculating the resulting stress tensors. The elastic constants were processed using the Voigt, Reuss, and Hill averaging schemes \cite{Zuo:gl0256,10.1063/1.1709944}, enabling the evaluation of directional Young’s moduli, shear moduli, and Poisson’s ratios.

Finally, optical properties including the absorption coefficient, reflectivity, and refractive index, were derived from the frequency-dependent complex dielectric function, calculated within the independent-particle approximation as implemented in CASTEP, following the methodology described in reference~\cite{lima2023dft}.

\section{Results}

\begin{figure}[!htb]
    \centering
    \includegraphics[width=\linewidth]{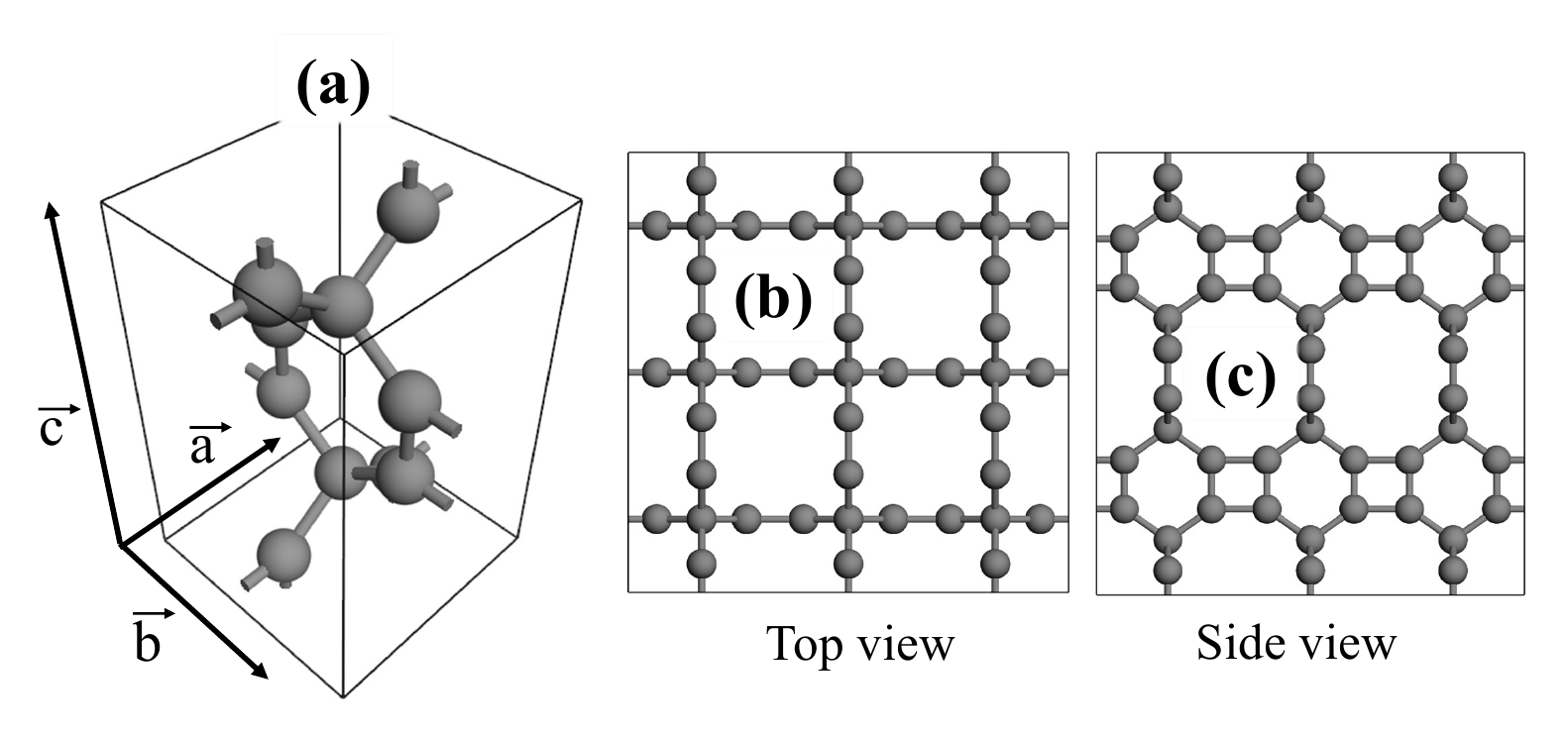}
    \caption{Optimized crystal structure of 3D-DPhyne. (a) Conventional unit cell highlighting the tetragonal lattice vectors $\vec{a}$, $\vec{b}$, and $\vec{c}$. (b) Top view along the $c$ axis, showing the nearly square in-plane projection and multiring arrangement. (c) Side view revealing the 3D covalent connectivity and the alternation of planar and nonplanar ring motifs. The structure consists of an interconnected network of four-, six-, and twelve-membered carbon rings with mixed sp/sp$^{2}$ hybridization, crystallizing in the $P4_{2}/mmc$ (No.~131) space group.}
    \label{fig:structure}
\end{figure}

Figure~\ref{fig:structure} shows the optimized crystal structure of 3D-DPhyne. The material crystallizes in a tetragonal lattice in the $P4_{2}/mmc$ (No.~131) space group, with optimized lattice parameters $a=b=3.94$~\AA{} and $c=6.11$~\AA{}. After full relaxation, the absence of residual forces and stresses confirms mechanical stability. Symmetry analysis reveals negligible deviations within 0.1~\AA{}, indicating a robust crystalline arrangement.

Structurally, 3D-DPhyne consists of interconnected four-, six-, and twelve-membered carbon rings, forming a non-honeycomb, multiring topology extending periodically in all crystallographic directions. This arrangement results in a framework with mixed sp/sp$^{2}$ hybridization, as linear carbon linkages coexist with planar and slightly puckered ring motifs. Consequently, this bonding diversity stabilizes the 3D architecture and distinguishes 3D-DPhyne from conventional 3D carbon phases, which possess only pure sp$^{3}$ (diamond-like) \cite{tiwari2016magical,sheng2011t,ding2020new} or sp$^{2}$ (graphitic) \cite{geim2009graphene,xu2013graphene} bonding.

The top and side views in Fig.~\ref{fig:structure}(b,c) highlight the 3D connectivity of the lattice. The in-plane projection reveals a nearly square arrangement consistent with tetragonal symmetry. In contrast, the side view shows alternating flat and puckered ring units repeating periodically along the $c$ axis. This structural feature promotes out-of-plane electronic coupling and distinguishes 3D-DPhyne from layered carbon systems, where weak van der Waals forces typically govern interlayer interactions. Here, fully covalent interlayer bonding creates a genuinely 3D $\pi$-conjugated network.

Compared with the other 3D carbon allotropes \cite{matar2024novel,matar2024novel,ju2025computational,fan2022four,yu2024novel,zhou2021oi20,yang2024explorative,gai2024two,li2015ab,fan2018d,katin2024diamanes}, 3D-DPhyne occupies an intermediate structural regime. Unlike the high symmetry and isotropic mechanical behavior of diamond and related sp$^{3}$-bonded phases \cite{lavini2022two}, 3D-DPhyne has a lower-symmetry lattice and greater topological complexity. In contrast to the layered stacking observed in graphitic \cite{harris2005new} or turbostratic \cite{li2007x} carbons, 3D-DPhyne features multiring motifs that provide additional degrees of freedom for electronic dispersion and elastic response. Unlike previously proposed porous 3D carbon frameworks, such as T-carbon \cite{sheng2011t} or other low-density open frameworks \cite{xu2013graphene}, 3D-DPhyne maintains a dense covalent backbone. It incorporates large ring units, thereby achieving structural anisotropy while retaining stability. The calculated formation energy of $-7.87$~eV/atom places 3D-DPhyne within the energy range of stable carbon allotropes \cite{shin2014cohesion}. This value matches those for several predicted 3D carbon phases \cite{matar2024novel,matar2024novel,ju2025computational,fan2022four,yu2024novel,zhou2021oi20,yang2024explorative,gai2024two,li2015ab,fan2018d,katin2024diamanes}.

\begin{table}[!htb]
\centering
\caption{Comparison between representative 3D carbon allotropes and 3D-DPhyne.}
\begin{tabular}{lcccc}
\hline
Allotrope & Hybridization & Symmetry & Topology & Electronic character \\
\hline
Diamond \cite{wort2008diamond} & sp$^{3}$ & Cubic & Tetrahedral network & Insulator \\
Graphite \cite{harris2005new} & sp$^{2}$ & Hexagonal & Layered honeycomb & Semimetal \\
T-carbon \cite{sheng2011t} & sp$^{3}$ & Cubic & Low-density porous & Semiconductor \\
3D graphyne-based phases \cite{liu2024graphyne} & sp/sp$^{2}$ & Various & Linear–ring networks & Metallic/Semiconducting \\
3D-DPhyne (this work) & sp/sp$^{2}$ & Tetragonal & 4–6–12 multiring network & Metallic \\
\hline
\end{tabular}
\end{table}

\begin{figure}[!htb]
    \centering
    \includegraphics[width=\linewidth]{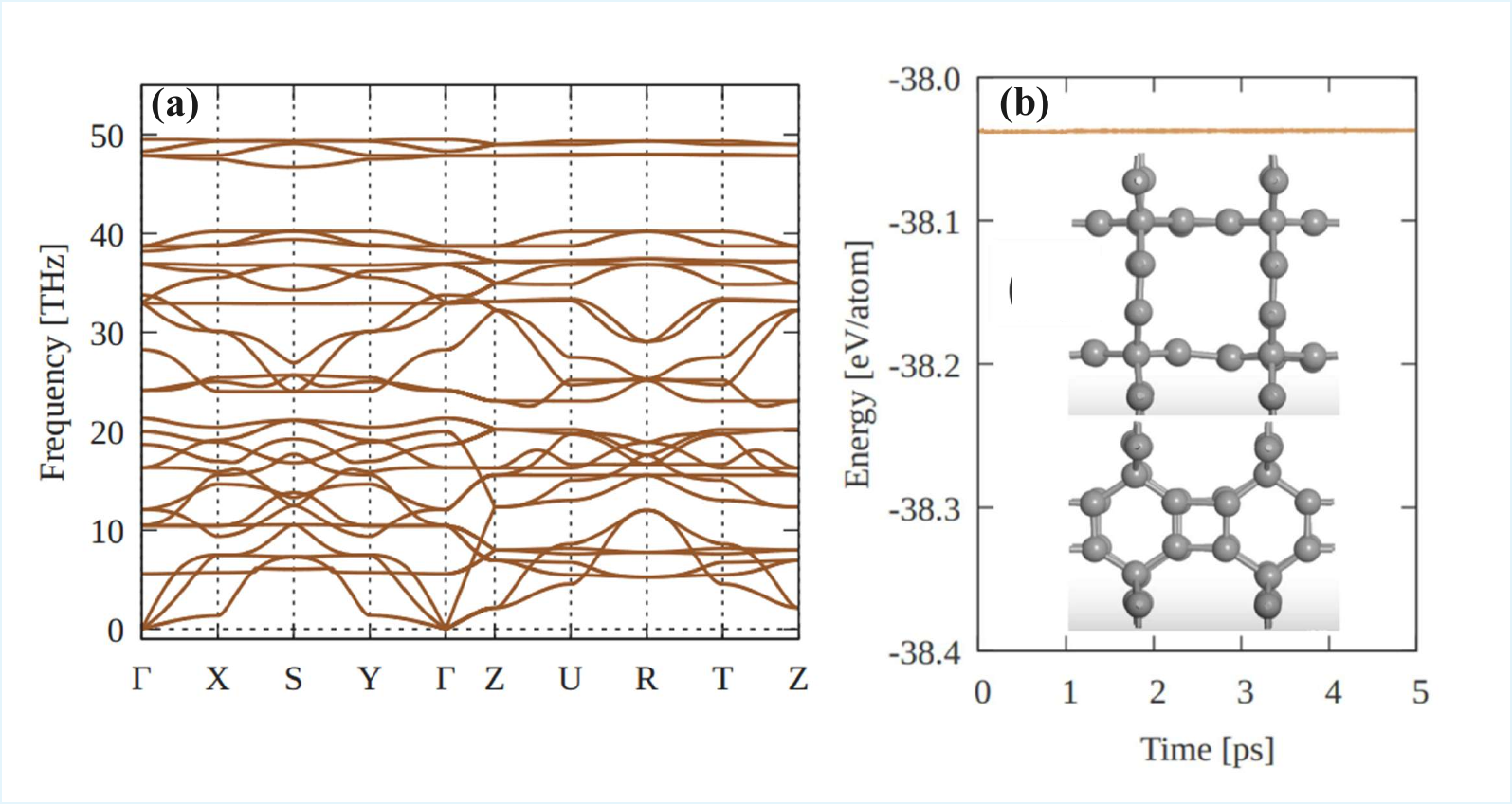}
    \caption{Dynamical and thermal stability of 3D-DPhyne. (a) Phonon dispersion relations calculated along high-symmetry directions of the Brillouin zone. (b) Time evolution of the total energy per atom obtained from AIMD simulations at 1000~K in the NVT ensemble over 5~ps. The inset shows the final atomic configuration.}
    \label{fig:phonon_aimd}
\end{figure}

The dynamical and thermal stability of 3D-DPhyne was assessed through phonon dispersion calculations and AIMD simulations, as shown in Fig.~\ref{fig:phonon_aimd}. The phonon spectrum along the high-symmetry directions of the Brillouin zone exhibits no imaginary frequencies, providing clear evidence that the optimized structure corresponds to a dynamically stable configuration. The acoustic branches display the expected linear behavior in the vicinity of the $\Gamma$ point, indicating mechanically stable long-wavelength lattice vibrations and well-defined elastic responses.

The overall phonon bandwidth reflects the coexistence of distinct bonding environments within the lattice. Lower-frequency modes are primarily associated with collective vibrations of the larger ring units and framework deformations. In contrast, higher-frequency optical modes originate from localized stretching vibrations involving sp and sp$^{2}$ carbon bonds. This separation of vibrational regimes is characteristic of multiring carbon networks. It has also been observed in other complex carbon allotropes that feature mixed hybridization and non-honeycomb topologies \cite{jiang2013r3,yang2024explorative}. Compared to purely sp$^{3}$-bonded phases such as diamond, which exhibit narrower vibrational distributions dominated by strong covalent bonds \cite{wort2008diamond}, 3D-DPhyne shows a richer phonon landscape. This arises from its topological complexity.

\begin{figure}[!htb]
    \centering
    \includegraphics[width=\linewidth]{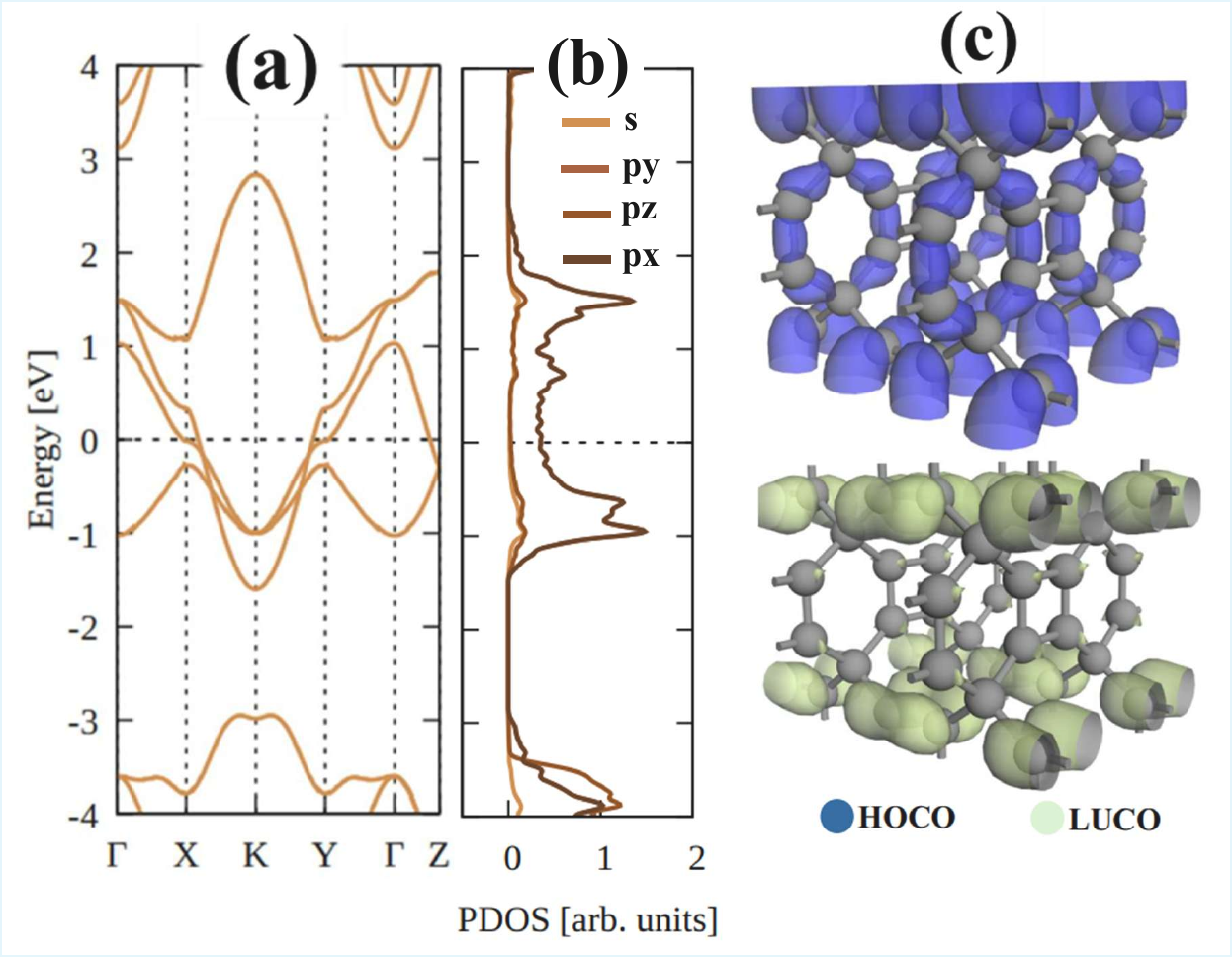}
    \caption{Electronic properties of 3D-DPhyne. (a) Electronic band structure calculated along high-symmetry directions of the Brillouin zone. The Fermi level is set to zero energy. (b) Projected density of states (PDOS), highlighting the dominant contribution of carbon $p$ orbitals near the Fermi level. (c) Real-space isosurfaces of the highest occupied crystal orbital (HOCO) and the lowest unoccupied crystal orbital (LUCO).}
    \label{fig:bands_pdos}
\end{figure}

To further test the thermal robustness of the proposed structure, we performed AIMD simulations at 1000 K in the NVT ensemble. Figure~\ref{fig:phonon_aimd}(b) shows that the total energy per atom stays nearly constant during the 5 ps simulation, with only minor thermal fluctuations and no abrupt drift. Structural snapshots at the end confirm that the crystal framework remains intact, there is no bond breaking, ring reconstruction, or loss of periodicity. This thermal resilience matches that of other metastable three-dimensional carbon allotropes and far exceeds the typical stability of low-coordination or weakly bonded carbon networks.

We now turn to the electronic properties of 3D-DPhyne, which were investigated through band structure calculations, projected density of states (PDOS), and real-space visualization of the frontier crystal orbitals, as shown in Fig.~\ref{fig:bands_pdos}. The calculated band structure reveals multiple bands crossing the Fermi level along the high-symmetry directions of the Brillouin zone, establishing the metallic character of the proposed 3D lattice. This behavior contrasts with that of its 2D precursor, dodecaphenylyne, which has been reported to be a narrow-gap semiconductor \cite{lima20252Ddodeca}.

\begin{figure}[!htb]
    \centering
    \includegraphics[width=0.7\linewidth]{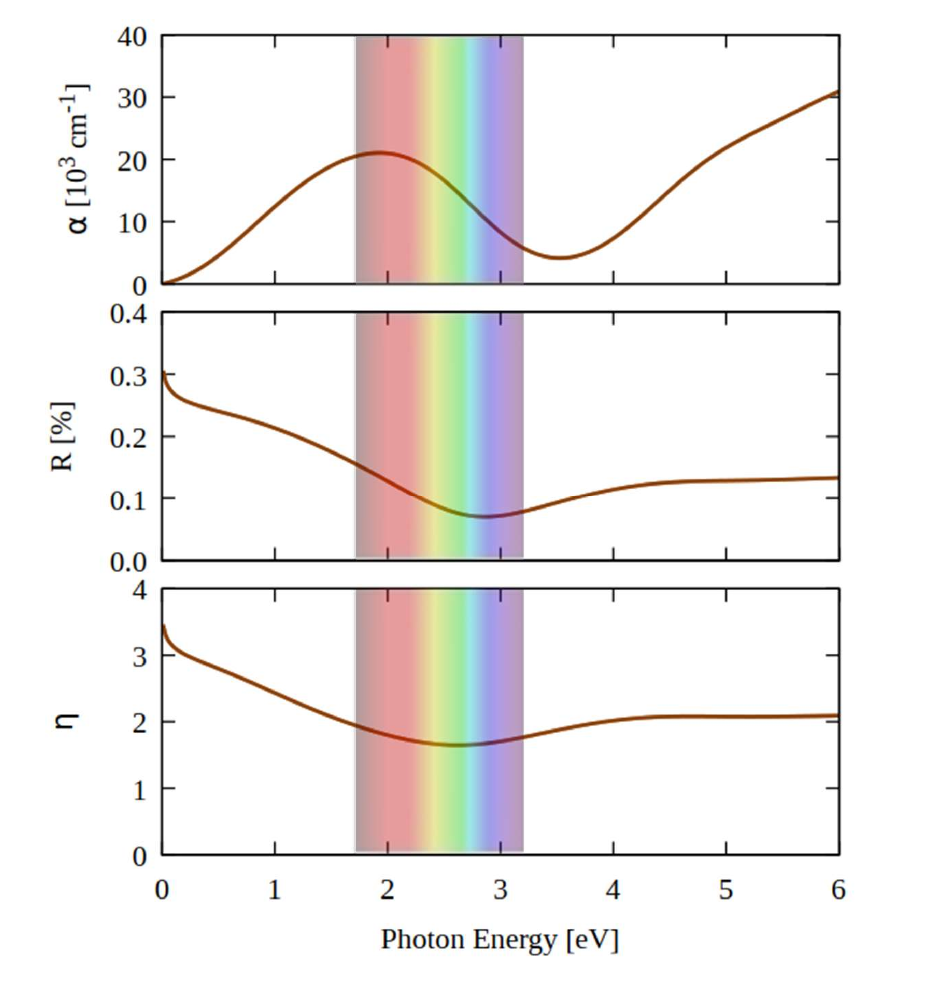}
    \caption{Optical properties of 3D-DPhyne obtained from the frequency-dependent complex dielectric function. (a) Absorption coefficient $\alpha(\omega)$, (b) Reflectivity $R(\omega)$, and (c) Refractive index $\eta(\omega)$ as a function of photon energy. The shaded area indicates the visible energy region.}
    \label{fig:optical}
\end{figure}

The metallicity observed in 3D-DPhyne can be directly linked to the mechanism of continuous electronic dispersion along the out-of-plane direction, arising from its fully covalent 3D connectivity. In layered carbon systems such as graphite, the weak interlayer coupling maintains a quasi-two-dimensional electronic character \cite{harris2005new}. In contrast, the extensive covalent bonding in 3D-DPhyne enables significant orbital overlap in all crystallographic directions, increasing band dispersion and closing the electronic gap. Thus, the dimensional crossover, driven by improved coordination and connectivity, induces delocalization of electronic states as observed in other 3D carbon networks derived from low-dimensional precursors \cite{felix20253d,tromer2024transforming}.

The PDOS analysis, shown in Fig.~\ref{fig:bands_pdos}(b), indicates that the electronic states near the Fermi level are dominated by carbon $p$ orbitals, with negligible contribution from $s$ states. This feature is characteristic of conjugated carbon systems and reflects the prevalence of $\pi$-type bonding throughout the lattice. In contrast to diamond, where sp$^{3}$ hybridization results in a wide band gap \cite{wort2008diamond}, or T-carbon \cite{sheng2011t}, which exhibits semiconducting behavior due to its tetrahedral bonding motif, 3D-DPhyne preserves a $\pi$-electron framework. It thus supports metallic conduction despite its 3D nature.

Further insight into the nature of the electronic states is provided by the real-space isosurfaces of the highest occupied crystal orbital (HOCO, blue) and the lowest unoccupied crystal orbital (LUCO, green), depicted in Fig.~\ref{fig:bands_pdos}(c). Both orbitals exhibit pronounced spatial delocalization across the entire 3D network, confirming the existence of an extended $\pi$-conjugated system. The LUCO exhibits enhanced interlayer connectivity relative to the HOCO, suggesting anisotropic electronic coupling that may contribute to direction-dependent transport properties. This behavior is distinct from that of conventional 3D carbon allotropes \cite{matar2024novel,matar2024novel,ju2025computational,fan2022four,yu2024novel,zhou2021oi20,yang2024explorative,gai2024two,li2015ab,fan2018d,katin2024diamanes} and underscores the role of multiring topology in shaping the electronic response.

The optical response of 3D-DPhyne was analyzed through the frequency-dependent absorption coefficient, reflectivity, and refractive index, as shown in Fig.~\ref{fig:optical}. These quantities were derived from the complex dielectric function within the independent-particle approximation \cite{lima2023dft} and provide insight into the interaction between electromagnetic radiation and the delocalized electronic states of the three-dimensional carbon framework.

Figure~\ref{fig:optical}(a) shows the absorption coefficient $\alpha(\omega)$, which demonstrates strong absorption across the visible and ultraviolet energy ranges. Specifically, in the visible region, $\alpha(\omega)$ reaches values around $10^{4}$~cm$^{-1}$ and increases at higher photon energies in the ultraviolet. This trend reflects the metallic electronic structure of 3D-DPhyne: low-energy interband transitions and intraband contributions from delocalized $\pi$ electrons give rise to broadband absorption. In contrast, wide-gap sp$^{3}$-bonded carbon phases such as diamond absorb little in the visible \cite{papadopoulos1991optical}, whereas 3D-DPhyne exhibits much greater optical activity. Graphite, whose optical absorption is strongly anisotropic and dominated by in-plane transitions \cite{djurivsic1999optical}, differs from 3D-DPhyne, in which 3D covalent connectivity yields a more uniform absorption response across crystallographic directions.

The reflectivity spectrum $R(\omega)$, shown in Fig.~\ref{fig:optical}(b), remains low over a broad energy window, with a minimum below 0.1 in the visible region. Such relatively low reflectivity is notable for a metallic carbon framework and can be attributed to the combination of a moderate carrier density and the absence of closely packed atomic planes, which are typically responsible for strong reflectivity in conventional metals. A similar reduction in reflectivity has been reported for other low-density or topologically complex 3D carbon allotropes \cite{felix2025novel,tromer2025electronic,felix20253d}, including porous and graphyne-derived networks \cite{hou2018study,kang2011elastic,gesesse2020analysis}, where electronic delocalization coexists with reduced optical screening.

The refractive index $\eta(\omega)$ in Fig.~\ref{fig:optical}(c) starts high at low photon energies, then gradually drops and stabilizes in the visible range. This trend shows the strong polarizability of the delocalized $\pi$-electron system and typifies carbon-based materials with extended conjugation. Unlike sp$^{3}$ carbon phases, which usually have lower refractive indices due to localized bonding \cite{fanchini2002optical}, 3D-DPhyne has greater optical polarizability because of its mixed sp/sp$^{2}$ hybridization and multiring topology.

Finally, the mechanical behavior of 3D-DPhyne was investigated by calculating its elastic constants and derived moduli, providing insight into its stiffness, deformation mechanisms, and anisotropy. The averaged bulk modulus ($K$), shear modulus ($G$), Young's modulus ($Y$), and Poisson's ratio ($\nu$), obtained using the Voigt, Reuss, and Hill schemes \cite{ziati2021mechanical,zuo1992elastic}, are summarized in Table~\ref{tab:elastic_avg}. The Hill averages yield $K = 159.09$~GPa, $Y = 161.79$~GPa, and $G = 60.80$~GPa, indicating a mechanically robust yet moderately compliant carbon framework.

\begin{table}[!htb]
\centering
\caption{Average elastic properties of 3D-DPhyne obtained using different averaging schemes.}
\label{tab:elastic_avg}
\begin{tabular}{lcccc}
\hline
Averaging scheme & $K$ (GPa) & $Y$ (GPa) & $G$ (GPa) & $\nu$ \\
\hline
Voigt  & 163.68 & 231.84 & 91.71 & 0.264 \\
Reuss  & 154.50 &  84.23 & 29.89 & 0.409 \\
Hill   & 159.09 & 161.79 & 60.80 & 0.331 \\
\hline
\end{tabular}
\end{table}

These values place 3D-DPhyne in an intermediate regime among 3D carbon allotropes. In contrast to diamond, which exhibits extremely high stiffness with Young's moduli exceeding 1~TPa due to its fully sp$^{3}$-bonded tetrahedral network, 3D-DPhyne shows reduced elastic rigidity arising from its mixed sp/sp$^{2}$ hybridization and multiring topology. On the other hand, its elastic moduli are significantly higher than those reported for low-density or highly porous carbon phases \cite{kabylda2025mechanical}, such as T-carbon \cite{sheng2011t,wang2019effects,bai2018mechanical} and related open-framework allotropes \cite{mortazavi2019prediction}, which typically display much lower resistance to deformation. This balance reflects the presence of a continuous covalent backbone, combined with structural motifs that enhance flexibility.

A key feature of the mechanical response of 3D-DPhyne is its pronounced elastic anisotropy, as evidenced by the wide ranges of directional Young's modulus, shear modulus, and Poisson's ratio reported in Table~\ref{tab:elastic_aniso}. Young's modulus varies from approximately 39 GPa to nearly 492 GPa, indicating strong direction-dependent stiffness associated with the underlying tetragonal symmetry and anisotropic connectivity of the multiring network. Such behavior contrasts sharply with highly symmetric cubic carbon phases, in which elastic properties are nearly isotropic, and highlights the critical role of topology in governing the mechanical response.

\begin{table}[!htb]
\centering
\caption{Directional variation of elastic moduli and Poisson’s ratio for 3D-DPhyne, highlighting elastic anisotropy.}
\label{tab:elastic_aniso}
\begin{tabular}{lcccccc}
\hline
Property & $E_{\mathrm{min}}$ (GPa) & $E_{\mathrm{max}}$ (GPa) & 
$G_{\mathrm{min}}$ (GPa) & $G_{\mathrm{max}}$ (GPa) & 
$\nu_{\mathrm{min}}$ & $\nu_{\mathrm{max}}$ \\
\hline
Value & 39.35 & 491.71 & 10.39 & 225.25 & $-0.102$ & 0.893 \\
\hline
Anisotropy & \multicolumn{2}{c}{12.49} & \multicolumn{2}{c}{21.67} & \multicolumn{2}{c}{--8.78} \\
\hline
\end{tabular}
\end{table}

The significant variation in Poisson’s ratio, including values exceeding conventional isotropic limits and negative values in specific directions, suggests unconventional deformation mechanisms enabled by the flexible ring-based architecture. Similar anomalous elastic responses have been predicted in other topologically complex carbon allotropes and metamaterial-like networks, where nonuniform bond rotation and ring deformation accommodate applied strain. In 3D-DPhyne, the coexistence of linear sp carbon linkages and planar sp$^{2}$ rings provides multiple deformation pathways, contributing to its unusual elastic behavior.

\begin{figure}[!htb]
    \centering
    \includegraphics[width=\linewidth]{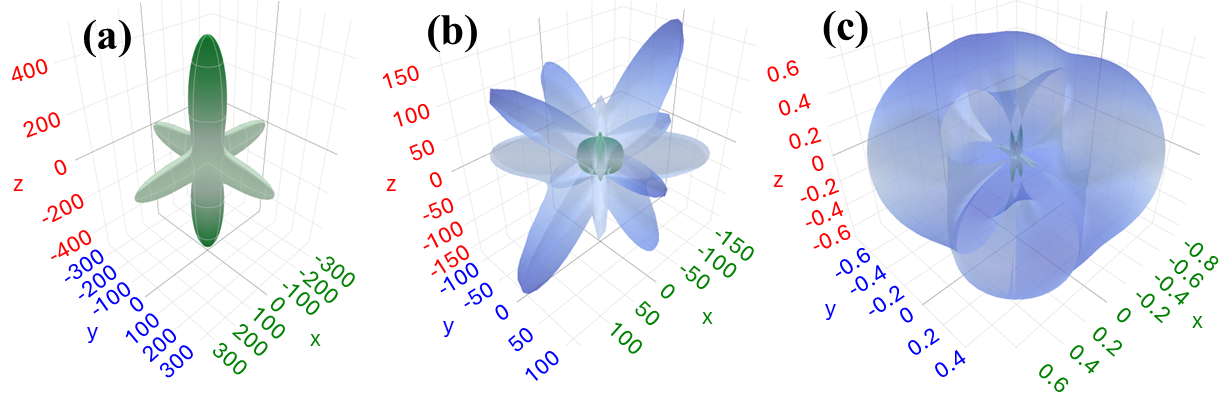}
    \caption{Three-dimensional directional dependence of the elastic properties of 3D-DPhyne. (a) Young's modulus $E(\mathbf{n})$, (b) shear modulus $G(\mathbf{n})$, and (c) Poisson's ratio $\nu(\mathbf{n})$ represented as polar surfaces.}
    \label{fig:elastic_surfaces}
\end{figure}

The pronounced elastic anisotropy of 3D-DPhyne is further illustrated in Fig.~\ref{fig:elastic_surfaces}, which presents the 3D directional dependence of the Young's modulus, shear modulus, and Poisson's ratio. These surfaces provide a comprehensive visualization of how the mechanical response varies with the orientation of the applied stress, offering deeper insight beyond averaged elastic constants.

Figure~\ref{fig:elastic_surfaces}(a) shows the directional Young's modulus $E(\mathbf{n})$. The distribution is highly non-spherical with elongated lobes along specific crystallographic directions. Maximum stiffness occurs along directions associated with continuous covalent pathways, formed by aligned ring motifs and linear carbon linkages. In contrast, much softer responses occur along directions where deformation is accommodated by ring bending and rotation. This behavior reflects the mixed sp/sp$^{2}$ hybridization and multiring topology of the lattice, which introduces direction-dependent load-bearing mechanisms. Highly symmetric 3D carbon phases, such as diamond, exhibit nearly isotropic elastic responses. In contrast, 3D-DPhyne displays stiffness variations exceeding one order of magnitude, consistent with the values reported in Table~\ref{tab:elastic_aniso}.

The shear modulus surface $G(\mathbf{n})$, shown in Fig.~\ref{fig:elastic_surfaces}(b), also exhibits strong anisotropy, with distinct lobes indicating preferred directions for shear resistance. Directions with high shear stiffness correspond to planes where covalent connectivity is maximized, whereas reduced shear resistance is observed along orientations dominated by flexible ring deformations. Such anisotropic shear behavior is characteristic of topologically complex frameworks and has been reported in other non-honeycomb carbon allotropes, although it remains uncommon among crystalline three-dimensional carbon phases.

The shear modulus surface $G(\mathbf{n})$, shown in Fig.~\ref{fig:elastic_surfaces}(b), exhibits pronounced anisotropy, with distinct lobes marking preferred directions for shear rigidity. In particular, directions with high shear stiffness align with planes where covalent bonding is most extensive, while diminished shear resistance arises along orientations governed by flexible ring motions. These anisotropic shear characteristics are characteristic of topologically complex frameworks and have been documented in other non-honeycomb carbon allotropes. However, they remain rare among crystalline 3D carbon phases.

Figure~\ref{fig:elastic_surfaces}(c) presents the directional Poisson's ratio $\nu(\mathbf{n})$, which displays an unconventional angular distribution, including regions of unusually high values and directions where negative Poisson's ratios emerge. This auxetic-like response arises from nonuniform deformation mechanisms enabled by the multiring architecture, in which applied strain can induce lateral expansion via cooperative bond rotation and ring rearrangement. Similar phenomena have been predicted in architected carbon networks and mechanical metamaterials \cite{saha2017coexistence,wang2019two,baimova2017auxetic,gao2021new}, but are notably absent in simple sp$^{3}$-bonded lattices.

Poisson's ratio values exceeding 0.5 in some crystallographic directions do not violate mechanical stability criteria. The classical upper bound of $\nu = 0.5$ applies only to isotropic, homogeneous, incompressible continua \cite{huang2016negative,dmitriev2001poisson}. In anisotropic crystals, especially those with low symmetry or complex topologies, Poisson's ratio varies with direction and can exceed this limit without signaling instability \cite{lethbridge2010elastic,grima2005auxetic}. In 3D-DPhyne, the multiring framework and mixed sp/sp$^{2}$ hybridization allow non-affine deformation mechanisms. Examples include bond rotation and ring distortion, which decouple transverse and longitudinal strains. Similar values are observed in architected solids \cite{greaves2011poisson}, auxetic metamaterials \cite{Grima.pssb.200460376}, and topologically complex carbon networks \cite{doi:10.1073/pnas.1717442115}, where compliance and lattice geometry govern elasticity. The large, anisotropic Poisson's ratios seen here are an intrinsic result of the lattice topology, not a calculation artifact.

\section{Conclusion}

In summary, we report a novel 3D carbon allotrope, denoted 3D-DPhyne, derived from the dodecaphenylyne framework. Using first-principles calculations, we have systematically characterized this new material. The optimized structure crystallizes in a tetragonal symmetry and consists of a topologically complex network of four-, six-, and twelve-membered carbon rings with mixed sp/sp$^{2}$ hybridization. Phonon dispersion relations and AIMD simulations confirm that the proposed framework is both dynamically and thermally stable, establishing its viability as a 3D carbon phase.

Electronic structure calculations show that expanding from a 2D dodecaphenylyne lattice to a fully covalent 3D framework induces metallic electronic character. This trend is supported by a delocalized $\pi$-electron network that extends along all crystallographic directions. The optical response mirrors this structure. It exhibits broadband absorption in the visible and ultraviolet regions with relatively low reflectivity, consistent with a metallic carbon framework with anisotropic polarizability.

A central result of this work is the pronounced mechanical anisotropy of 3D-DPhyne. Directional elastic analyses show substantial variations in Young's and shear moduli, indicating that mechanical stiffness depends on crystallographic orientation. Poisson's ratio exhibits unconventional directional behavior, including values above the isotropic limit and auxetic-like responses in specific directions. This behavior arises from the multiring topology and non-affine deformation mechanisms enabled by the mixed-bonding network, distinguishing 3D-DPhyne from conventional sp$^{3}$-bonded and layered carbon phases.

\section*{Acknowledgments}
This work was supported by the Brazilian funding agencies Fundação de Amparo à Pesquisa do Estado de São Paulo (FAPESP) (grants no. 2022/03959-6, 2022/14576-0, 2013/08293-7, 2020/01144-0, 2024/05087-1, and 2022/16509-9), National Council for Scientific, Technological Development (CNPq) (grants no. 307213/2021–8, 350176/2022-1, and 167745/2023-9), FAP-DF (grants no. 00193.00001808/2022-71 and 00193-00001857/2023-95), FAPDF-PRONEM (grant no. 00193.00001247/2021-20), and PDPG-FAPDF-CAPES Centro-Oeste (grant no. 00193-00000867/2024-94). 

\bibliographystyle{unsrtnat}
\bibliography{references}  

@article{xiao2022dimensionality,
  title={Dimensionality, function and performance of carbon materials in energy storage devices},
  author={Xiao, Jing and Han, Junwei and Zhang, Chen and Ling, Guowei and Kang, Feiyu and Yang, Quan-Hong},
  journal={Advanced Energy Materials},
  volume={12},
  number={4},
  pages={2100775},
  year={2022},
  publisher={Wiley Online Library}
}

@article{speranza2019role,
  title={The role of functionalization in the applications of carbon materials: An overview},
  author={Speranza, Giorgio},
  journal={C},
  volume={5},
  number={4},
  pages={84},
  year={2019},
  publisher={MDPI}
}

@article{nasir2018carbon,
  title={Carbon-based nanomaterials/allotropes: A glimpse of their synthesis, properties and some applications},
  author={Nasir, Salisu and Hussein, Mohd Zobir and Zainal, Zulkarnain and Yusof, Nor Azah},
  journal={Materials},
  volume={11},
  number={2},
  pages={295},
  year={2018},
  publisher={MDPI}
}

@article{tiwari2016magical,
  title={Magical allotropes of carbon: prospects and applications},
  author={Tiwari, Santosh K and Kumar, Vijay and Huczko, Andrzej and Oraon, R and Adhikari, A De and Nayak, GC},
  journal={Critical Reviews in Solid State and Materials Sciences},
  volume={41},
  number={4},
  pages={257--317},
  year={2016},
  publisher={Taylor \& Francis}
}

@article{geim2009graphene,
  title={Graphene: status and prospects},
  author={Geim, Andre Konstantin},
  journal={science},
  volume={324},
  number={5934},
  pages={1530--1534},
  year={2009},
  publisher={American Association for the Advancement of Science}
}

@incollection{yadav2024carbon,
  title={Carbon allotropes: past to present aspects},
  author={Yadav, Chandra Shekhar and Azad, Iqbal and Khan, Abdul Rahman and Singh, Prashant},
  booktitle={Biosensors based on graphene, graphene oxide and graphynes for early detection of cancer},
  pages={1--23},
  year={2024},
  publisher={CRC Press}
}

@article{rani2013designing,
  title={Designing band gap of graphene by B and N dopant atoms},
  author={Rani, Pooja and Jindal, VK},
  journal={RSC advances},
  volume={3},
  number={3},
  pages={802--812},
  year={2013},
  publisher={Royal Society of Chemistry}
}

@article{enyashin2011graphene,
  title={Graphene allotropes},
  author={Enyashin, Andrey N and Ivanovskii, Alexander L},
  journal={physica status solidi (b)},
  volume={248},
  number={8},
  pages={1879--1883},
  year={2011},
  publisher={Wiley Online Library}
}

@article{lusk2009creation,
  title={Creation of graphene allotropes using patterned defects},
  author={Lusk, Mark T and Carr, LD},
  journal={Carbon},
  volume={47},
  number={9},
  pages={2226--2232},
  year={2009},
  publisher={Elsevier}
}

@article{fan2021biphenylene,
  title={Biphenylene network: A nonbenzenoid carbon allotrope},
  author={Fan, Qitang and Yan, Linghao and Tripp, Matthias W and Krej{\v{c}}{\'\i}, Ond{\v{r}}ej and Dimosthenous, Stavrina and Kachel, Stefan R and Chen, Mengyi and Foster, Adam S and Koert, Ulrich and Liljeroth, Peter and others},
  journal={Science},
  volume={372},
  number={6544},
  pages={852--856},
  year={2021},
  publisher={American Association for the Advancement of Science}
}

@article{hou2022synthesis,
  title={Synthesis of a monolayer fullerene network},
  author={Hou, Lingxiang and Cui, Xueping and Guan, Bo and Wang, Shaozhi and Li, Ruian and Liu, Yunqi and Zhu, Daoben and Zheng, Jian},
  journal={Nature},
  volume={606},
  number={7914},
  pages={507--510},
  year={2022},
  publisher={Nature Publishing Group UK London}
}

@article{zhang2019recent,
  title={Recent advances of porous graphene: synthesis, functionalization, and electrochemical applications},
  author={Zhang, Yuanyuan and Wan, Qijin and Yang, Nianjun},
  journal={Small},
  volume={15},
  number={48},
  pages={1903780},
  year={2019},
  publisher={Wiley Online Library}
}

@article{shehzad2016three,
  title={Three-dimensional macro-structures of two-dimensional nanomaterials},
  author={Shehzad, Khurram and Xu, Yang and Gao, Chao and Duan, Xiangfeng},
  journal={Chemical Society Reviews},
  volume={45},
  number={20},
  pages={5541--5588},
  year={2016},
  publisher={Royal Society of Chemistry}
}

@article{fu2017progress,
  title={Progress in 3D printing of carbon materials for energy-related applications},
  author={Fu, Kun and Yao, Yonggang and Dai, Jiaqi and Hu, Liangbing},
  journal={Advanced materials},
  volume={29},
  number={9},
  pages={1603486},
  year={2017},
  publisher={Wiley Online Library}
}

@article{ding2020new,
  title={A new carbon allotrope: T5-carbon},
  author={Ding, Xian-Yong and Zhang, Chao and Wang, Dong-Qi and Li, Bing-Sheng and Wang, Qingping and Yu, Zhi Gen and Ang, Kah-Wee and Zhang, Yong-Wei},
  journal={Scripta Materialia},
  volume={189},
  pages={72--77},
  year={2020},
  publisher={Elsevier}
}

@article{xu2013graphene,
  title={Graphene-like two-dimensional materials},
  author={Xu, Mingsheng and Liang, Tao and Shi, Minmin and Chen, Hongzheng},
  journal={Chemical reviews},
  volume={113},
  number={5},
  pages={3766--3798},
  year={2013},
  publisher={ACS Publications}
}

@article{sheng2011t,
  title={T-carbon: a novel carbon allotrope},
  author={Sheng, Xian-Lei and Yan, Qing-Bo and Ye, Fei and Zheng, Qing-Rong and Su, Gang},
  journal={Physical review letters},
  volume={106},
  number={15},
  pages={155703},
  year={2011},
  publisher={APS}
}

@article{liu2020carbon,
  title={Carbon foams: 3D porous carbon materials holding immense potential},
  author={Liu, Heguang and Wu, Shaoqing and Tian, Na and Yan, Fuxue and You, Caiyin and Yang, Yang},
  journal={Journal of Materials Chemistry A},
  volume={8},
  number={45},
  pages={23699--23723},
  year={2020},
  publisher={Royal Society of Chemistry}
}

@article{sun20203d,
  title={3D graphene materials: from understanding to design and synthesis control},
  author={Sun, Zhuxing and Fang, Siyuan and Hu, Yun Hang},
  journal={Chemical Reviews},
  volume={120},
  number={18},
  pages={10336--10453},
  year={2020},
  publisher={ACS Publications}
}

@article{LIMA2025117868,
title = {Athos-Graphene: Computational discovery of an art-inspired 2D carbon anode for lithium-ion batteries},
journal = {Journal of Energy Storage},
volume = {133},
pages = {117868},
year = {2025},
doi = {https://doi.org/10.1016/j.est.2025.117868},
author = {Kleuton A.L. Lima and José A.S. Laranjeira and Nicolas F. Martins and Julio R. Sambrano and Alexandre C. Dias and Douglas S. Galvão and Luiz A. Ribeiro Junior}
}

@article{lima2025structural,
  title={Structural, electronic, and Li-ion adsorption properties of PolyPyGY explored by first-principles and machine learning simulations: A new multi-ringed 2D carbon allotrope},
  author={Lima, KAL and da Silva, Daniel Alves and Nze, GD Amvame and de Mendon{\c{c}}a, FL Lopes and Pereira Jr, ML and Ribeiro Jr, LA},
  journal={Journal of Energy Storage},
  volume={117},
  pages={116099},
  year={2025},
  publisher={Elsevier}
}

@article{lima2025petal,
  title={Petal-Graphyne: A novel 2D carbon allotrope for high-performance Li and Na ion storage},
  author={Lima, Kleuton AL and Laranjeira, Jos{\'e} AS and Martins, Nicolas F and Dias, Alexandre C and Sambrano, Julio R and Galv{\~a}o, Douglas S and Junior, Luiz A Ribeiro},
  journal={Journal of Energy Storage},
  volume={130},
  pages={117235},
  year={2025},
  publisher={Elsevier}
}

@article{lima2025first,
  title={First-principles and machine learning insights into the design of DOTT-carbon and its lithium-ion storage capacity},
  author={Lima, Kleuton AL and Abreu, Ana VP and Silva, Alysson MA and Ribeiro, Luiz A},
  journal={Journal of Materials Chemistry A},
  volume={13},
  number={21},
  pages={15609--15619},
  year={2025},
  publisher={Royal Society of Chemistry}
}

@article{mortazavi2023electronic,
  title={Electronic, thermal and mechanical properties of carbon and boron nitride holey graphyne monolayers},
  author={Mortazavi, Bohayra},
  journal={Materials},
  volume={16},
  number={20},
  pages={6642},
  year={2023},
  publisher={MDPI}
}

@article{bafekry2024layered,
  title={Layered conjugated porous fused aromatic network structures of two-dimensional carbon nitride: a first-principles calculation of optoelectronic properties},
  author={Bafekry, A and Fadlallah, MM and Stampfl, C and Ziabari, A Abdolahzadeh and Fazeli, S and Faraji, M and Jappor, HR and Ghergherehchi, M},
  journal={Applied Physics A},
  volume={130},
  number={7},
  pages={500},
  year={2024},
  publisher={Springer}
}

@article{bafekry2022theoretical,
  title={Theoretical prediction of two-dimensional BC2X (X= N, P, As) monolayers: ab initio investigations},
  author={Bafekry, A and Naseri, M and Faraji, M and Fadlallah, MM and Hoat, DM and Jappor, HR and Ghergherehchi, M and Gogova, D and Afarideh, H},
  journal={Scientific Reports},
  volume={12},
  number={1},
  pages={22269},
  year={2022},
  publisher={Nature Publishing Group UK London}
}

@article{bafekry2022tunable,
  title={Tunable electronic properties of porous graphitic carbon nitride (C6N7) monolayer by atomic doping and embedding: A first-principle study},
  author={Bafekry, A and Faraji, M and Hieu, NN and Khatibani, A Bagheri and Fadlallah, Mohamed M and Gogova, D and Ghergherehchi, M},
  journal={Applied Surface Science},
  volume={583},
  pages={152270},
  year={2022},
  publisher={Elsevier}
}

@article{felix2025novel,
  title={Novel 3D Pentagraphene Allotropes: Stability, Electronic, Mechanical, and Optical Properties},
  author={F{\'e}lix, IM and Ipaves, B and de Oliveira, RB and Junior, LA and Rocha, LS and Junior, ML and Galv{\~a}o, DS and Tromer, RM},
  journal={arXiv preprint arXiv:2509.10191},
  year={2025}
}

@article{tromer2025electronic,
  title={On the Electronic, Mechanical and Optical Properties of Superhard Cross-Linked Carbon Nanotubes (Tubulanes)},
  author={Tromer, Raphael M and Ipaves, Bruno and Junior, Marcelo L Pereira and Woellner, Cristiano F and Cai, Kun and Galvao, Douglas S},
  journal={arXiv preprint arXiv:2509.09571},
  year={2025}
}

@article{felix20253d,
  title={3D Carbon Nanostructures Derived from 2D Irida-Graphene: Insights into Structural, Mechanical, Electronic, and Optical Properties},
  author={Felix, Isaac M and de Oliveira, Raphael B and Ribeiro Jr, Luiz A and Galv{\~a}o, Douglas S and Pereira Jr, Marcelo L and Tromer, Raphael M},
  journal={ACS omega},
  volume={10},
  number={34},
  pages={38985--38994},
  year={2025},
  publisher={ACS Publications}
}

@article{tromer2024transforming,
  title={Transforming Two-Dimensional Carbon Allotropes into Three-Dimensional Ones through Topological Mapping: The Case of Biphenylene Carbon (Graphenylene)},
  author={Tromer, Raphael M and Felix, Levi C and Baughmann, Ray H and Galvao, Douglas S and Woellner, Cristiano F},
  journal={The Journal of Physical Chemistry A},
  volume={128},
  number={35},
  pages={7346--7352},
  year={2024},
  publisher={ACS Publications}
}

@article{lebedeva2007kinetics,
  title={The kinetics of carbon nanostructure 2D--3D transformation},
  author={Lebedeva, IV and Knizhnik, AA and Potapkin, BV},
  journal={Russian Journal of Physical Chemistry B},
  volume={1},
  number={6},
  pages={675--684},
  year={2007},
  publisher={Springer}
}

@article{fayos1999possible,
  title={Possible 3D carbon structures as progressive intermediates in graphite to diamond phase transition},
  author={Fayos, Jos{\'e}},
  journal={Journal of Solid State Chemistry},
  volume={148},
  number={2},
  pages={278--285},
  year={1999},
  publisher={Elsevier}
}

@article{lima2026first,
  title={First-principles and machine learning investigation of the structural and optoelectronic properties of dodecaphenylyne: a novel carbon allotrope},
  author={Lima, Kleuton AL and Laranjeira, Jos{\'e} AS and Martins, Nicolas F and Sambrano, Julio R and Dias, Alexandre C and Junior, Luiz A Ribeiro and Galv{\~a}o, Douglas S},
  journal={Nanoscale},
  year={2026},
  publisher={Royal Society of Chemistry}
}

@article{aliev2025planar,
  title={A planar-sheet nongraphitic zero-bandgap sp2 carbon phase made by the low-temperature reaction of $\gamma$-graphyne},
  author={Aliev, Ali E and Guo, Yongzhe and Fonseca, Alexandre F and Razal, Joselito M and Wang, Zhong and Galv{\~a}o, Douglas S and Bolding, Claire M and Chapman-Wilson, Nathaniel E and Desyatkin, Victor G and Leisen, Johannes E and others},
  journal={Proceedings of the National Academy of Sciences},
  volume={122},
  number={5},
  pages={e2413194122},
  year={2025},
  publisher={National Academy of Sciences}
}

@article{desyatkin2022scalable,
  title={Scalable synthesis and characterization of multilayer $\gamma$-graphyne, new carbon crystals with a small direct band gap},
  author={Desyatkin, Victor G and Martin, William B and Aliev, Ali E and Chapman, Nathaniel E and Fonseca, Alexandre F and Galv{\~a}o, Douglas S and Miller, Ericka Roy and Stone, Kevin H and Wang, Zhong and Zakhidov, Dante and others},
  journal={Journal of the American Chemical Society},
  volume={144},
  number={39},
  pages={17999--18008},
  year={2022},
  publisher={ACS Publications}
}

@article{clark2005first,
  title={First principles methods using CASTEP},
  author={Clark, Stewart J and Segall, Matthew D and Pickard, Chris J and Hasnip, Phil J and Probert, Matt IJ and Refson, Keith and Payne, Mike C},
  journal={Zeitschrift f{\"u}r kristallographie-crystalline materials},
  volume={220},
  number={5-6},
  pages={567--570},
  year={2005},
  publisher={Oldenbourg Wissenschaftsverlag}
}

@article{perdew1996generalized,
  title={Generalized gradient approximation made simple},
  author={Perdew, John P and Burke, Kieron and Ernzerhof, Matthias},
  journal={Physical review letters},
  volume={77},
  number={18},
  pages={3865},
  year={1996},
  publisher={APS}
}

@article{head1985broyden,
  title={A Broyden—Fletcher—Goldfarb—Shanno optimization procedure for molecular geometries},
  author={Head, John D and Zerner, Michael C},
  journal={Chemical physics letters},
  volume={122},
  number={3},
  pages={264--270},
  year={1985},
  publisher={Elsevier}
}

@article{PFROMMER1997233,
  title = {Relaxation of Crystals with the Quasi-Newton Method},
  journal = {Journal of Computational Physics},
  volume = {131},
  number = {1},
  pages = {233-240}, 
  year = {1997},
  issn = {0021-9991},
  author = {Bernd G. Pfrommer and Michel Côté and Steven G. Louie and Marvin L. Cohen}
}

@article{Zuo:gl0256,
  author = "Zuo, L. and Humbert, M. and Esling, C.",
  title = "{Elastic properties of polycrystals in the Voigt-Reuss-Hill approximation}",
  journal = "Journal of Applied Crystallography",
  year = "1992",
  volume = "25",
  number = "6",
  pages = "751--755",
  month = "Dec"
}

@article{10.1063/1.1709944,
  author = {Chung, D. H. and Buessem, W. R.},
  title = "{The Voigt‐Reuss‐Hill Approximation and Elastic Moduli of Polycrystalline MgO, CaF2, $\beta$‐ZnS, ZnSe, and CdTe}",
  journal = {Journal of Applied Physics},
  volume = {38},
  number = {6},
  pages = {2535-2540},
  year = {2004},
  month = {06}
}

@article{lima2023dft,
  title={A DFT study on the mechanical, electronic, thermodynamic, and optical properties of GaN and AlN counterparts of biphenylene network},
  author={Lima, KA Lopes and Junior, LA Ribeiro},
  journal={Materials Today Communications},
  pages={107183},
  year={2023},
  publisher={Elsevier}
}

@article{matar2024novel,
  title={Novel Superhard Tetragonal Hybrid sp3/sp2 Carbon Allotropes C x (x= 5, 6, 7): Crystal Chemistry and Ab Initio Studies},
  author={Matar, Samir F and Solozhenko, Vladimir L},
  journal={C},
  volume={10},
  number={3},
  pages={64},
  year={2024},
  publisher={MDPI}
}

@article{ju2025computational,
  title={Computational Investigation of an All-sp 3 Hybridized Superstable Carbon Allotrope with Large Band Gap},
  author={Ju, Xiaoshi and Bu, Kun and Zhang, Chunxiao and Sun, Yuping},
  journal={Materials},
  volume={18},
  number={11},
  pages={2533},
  year={2025},
  publisher={MDPI}
}

@article{fan2022four,
  title={Four carbon allotropes form COT structures},
  author={Fan, Qingyang and Liu, Heng and Jiang, Li and Zhang, Wei and Yu, Xinhai and Yun, Sining},
  journal={ACS Applied Electronic Materials},
  volume={4},
  number={5},
  pages={2353--2363},
  year={2022},
  publisher={ACS Publications}
}

@article{yu2024novel,
  title={A novel sp 3 carbon allotrope with 4+ 5+ 6+ 7+ 8 odd--even ring},
  author={Yu, Tao and Zhu, Sheng-Cai and Hou, Yanglong},
  journal={Physical Chemistry Chemical Physics},
  volume={26},
  number={33},
  pages={22182--22188},
  year={2024},
  publisher={Royal Society of Chemistry}
}

@article{zhou2021oi20,
  title={oI20-carbon: A new superhard carbon allotrope},
  author={Zhou, Lin and Chai, Changchun and Zhang, Wei and Song, Yanxing and Zhang, Zheren and Yang, Yintang},
  journal={Diamond and Related Materials},
  volume={113},
  pages={108284},
  year={2021},
  publisher={Elsevier}
}

@article{yang2024explorative,
  title={Explorative prediction of novel superhard carbon allotropes with lager cell: Density functional theory-assisted deep learning},
  author={Yang, Jiangtao and Fan, Qingyang and Ye, Ming and Liu, Heng},
  journal={Diamond and Related Materials},
  volume={147},
  pages={111320},
  year={2024},
  publisher={Elsevier}
}

@article{gai2024two,
  title={Two novel carbon allotropes with exceptional properties as superhard materials},
  author={Gai, Xiaoqian and Tian, Fubo and Cui, Tian and Yang, Mengxin},
  journal={Diamond and Related Materials},
  volume={148},
  pages={111492},
  year={2024},
  publisher={Elsevier}
}

@article{li2015ab,
  title={Ab initio structure determination of n-diamond},
  author={Li, Da and Tian, Fubo and Chu, Binhua and Duan, Defang and Sha, Xiaojing and Lv, Yunzhou and Zhang, Huadi and Lu, Nan and Liu, Bingbing and Cui, Tian},
  journal={Scientific Reports},
  volume={5},
  number={1},
  pages={13447},
  year={2015},
  publisher={Nature Publishing Group UK London}
}

@article{fan2018d,
  title={D-carbon: Ab initio study of a novel carbon allotrope},
  author={Fan, Dong and Lu, Shaohua and Golov, Andrey A and Kabanov, Artem A and Hu, Xiaojun},
  journal={The Journal of chemical physics},
  volume={149},
  number={11},
  year={2018},
  publisher={AIP Publishing}
}

@article{lavini2022two,
  title={Two-dimensional diamonds from sp 2-to-sp 3 phase transitions},
  author={Lavini, Francesco and Rejhon, Martin and Riedo, Elisa},
  journal={Nature Reviews Materials},
  volume={7},
  number={10},
  pages={814--832},
  year={2022},
  publisher={Nature Publishing Group UK London}
}

@article{harris2005new,
  title={New perspectives on the structure of graphitic carbons},
  author={Harris, Peter JF},
  journal={Critical reviews in solid state and materials sciences},
  volume={30},
  number={4},
  pages={235--253},
  year={2005},
  publisher={Taylor \& Francis}
}

@article{li2007x,
  title={X-ray diffraction patterns of graphite and turbostratic carbon},
  author={Li, ZQ and Lu, CJ and Xia, ZP and Zhou, Y and Luo, Z},
  journal={Carbon},
  volume={45},
  number={8},
  pages={1686--1695},
  year={2007},
  publisher={Elsevier}
}

@article{shin2014cohesion,
  title={Cohesion energetics of carbon allotropes: Quantum Monte Carlo study},
  author={Shin, Hyeondeok and Kang, Sinabro and Koo, Jahyun and Lee, Hoonkyung and Kim, Jeongnim and Kwon, Yongkyung},
  journal={The Journal of chemical physics},
  volume={140},
  number={11},
  year={2014},
  publisher={AIP Publishing}
}

@article{katin2024diamanes,
  title={Diamanes from novel graphene allotropes: Computational study on structures, stabilities and properties},
  author={Katin, Konstantin P and Podlivaev, Alexey I and Kochaev, Alexei I and Kulyamin, Pavel A and Bauetdinov, Yusupbek and Grekova, Anastasiya A and Bereznitskiy, Igor V and Maslov, Mikhail M},
  journal={FlatChem},
  volume={44},
  pages={100622},
  year={2024},
  publisher={Elsevier}
}

@article{wort2008diamond,
  title={Diamond as an electronic material},
  author={Wort, Chris JH and Balmer, Richard S},
  journal={Materials today},
  volume={11},
  number={1-2},
  pages={22--28},
  year={2008},
  publisher={Elsevier}
}

@article{liu2024graphyne,
  title={Graphyne-based 3D porous structure and its sandwich-type graphene structure for alkali metal ion battery anode materials},
  author={Liu, Haidong and Li, Xiaowei and Chen, Haotian and Chen, Jin and Shi, Zixun},
  journal={Physical Chemistry Chemical Physics},
  volume={26},
  number={10},
  pages={8426--8435},
  year={2024},
  publisher={Royal Society of Chemistry}
}

@article{papadopoulos1991optical,
  title={Optical properties of diamond},
  author={Papadopoulos, AD and Anastassakis, E},
  journal={Physical Review B},
  volume={43},
  number={6},
  pages={5090},
  year={1991},
  publisher={APS}
}

@article{djurivsic1999optical,
  title={Optical properties of graphite},
  author={Djuri{\v{s}}i{\'c}, Aleksandra B and Li, E Herbert},
  journal={Journal of applied physics},
  volume={85},
  number={10},
  pages={7404--7410},
  year={1999},
  publisher={American Institute of Physics}
}

@article{hou2018study,
  title={Study of electronic structure, thermal conductivity, elastic and optical properties of $\alpha$, $\beta$, $\gamma$-graphyne},
  author={Hou, Xun and Xie, Zhongjing and Li, Chunmei and Li, Guannan and Chen, Zhiqian},
  journal={Materials},
  volume={11},
  number={2},
  pages={188},
  year={2018},
  publisher={MDPI}
}

@article{kang2011elastic,
  title={Elastic, electronic, and optical properties of two-dimensional graphyne sheet},
  author={Kang, Jun and Li, Jingbo and Wu, Fengmin and Li, Shu-Shen and Xia, Jian-Bai},
  journal={The Journal of Physical Chemistry C},
  volume={115},
  number={42},
  pages={20466--20470},
  year={2011},
  publisher={ACS Publications}
}

@article{gesesse2020analysis,
  title={On the analysis of diffuse reflectance measurements to estimate the optical properties of amorphous porous carbons and semiconductor/carbon catalysts},
  author={Gesesse, Getaneh Diress and Gomis-Berenguer, Alicia and Barthe, Marie-France and Ania, Conchi O},
  journal={Journal of Photochemistry and Photobiology A: Chemistry},
  volume={398},
  pages={112622},
  year={2020},
  publisher={Elsevier}
}

@article{fanchini2002optical,
  title={Optical properties of disordered carbon-based materials},
  author={Fanchini, G and Ray, SC and Tagliaferro, Alberto},
  journal={Surface and Coatings Technology},
  volume={151},
  pages={233--241},
  year={2002},
  publisher={Elsevier}
}

@article{ziati2021mechanical,
  title={Mechanical properties and thermodynamic parameters of Sr 2 RuO 4 and Sr 2 RuO 2 F 2 compounds under pressure and temperature effects: Voigt--Reuss--Hill approximations and Debye model},
  author={Ziati, Meryem and Ez-Zahraouy, Hamid},
  journal={J. Phys. Opt. Sci},
  volume={3},
  pages={2--8},
  year={2021}
}

@article{zuo1992elastic,
  title={Elastic properties of polycrystals in the Voigt-Reuss-Hill approximation},
  author={Zuo, LIANG and Humbert, MICHEL and Esling, CLAUDE},
  journal={Applied Crystallography},
  volume={25},
  number={6},
  pages={751--755},
  year={1992},
  publisher={International Union of Crystallography}
}

@article{mortazavi2019prediction,
  title={Prediction of C 7 N 6 and C 9 N 4: stable and strong porous carbon-nitride nanosheets with attractive electronic and optical properties},
  author={Mortazavi, Bohayra and Shahrokhi, Masoud and Shapeev, Alexander V and Rabczuk, Timon and Zhuang, Xiaoying},
  journal={Journal of Materials Chemistry C},
  volume={7},
  number={35},
  pages={10908--10917},
  year={2019},
  publisher={Royal Society of Chemistry}
}

@article{kabylda2025mechanical,
  title={Mechanical Properties of Nanoporous Graphenes: Transferability of Graph Machine-Learned Force Fields Compared to Local and Reactive Potentials},
  author={Kabylda, Adil and Mortazavi, Bohayra and Zhuang, Xiaoying and Tkatchenko, Alexandre},
  journal={Advanced Functional Materials},
  volume={35},
  number={13},
  pages={2417891},
  year={2025},
  publisher={Wiley Online Library}
}

@article{wang2019effects,
  title={Effects of tensile strain rate and grain size on the mechanical properties of nanocrystalline T-carbon},
  author={Wang, Ying and Lei, Jincheng and Bai, Lichun and Zhou, Kun and Liu, Zishun},
  journal={Computational Materials Science},
  volume={170},
  pages={109188},
  year={2019},
  publisher={Elsevier}
}

@article{bai2018mechanical,
  title={Mechanical behaviors of T-carbon: A molecular dynamics study},
  author={Bai, Lichun and Sun, Ping-Ping and Liu, Bo and Liu, Zishun and Zhou, Kun},
  journal={Carbon},
  volume={138},
  pages={357--362},
  year={2018},
  publisher={Elsevier}
}

@article{saha2017coexistence,
  title={Coexistence of normal and auxetic behavior in a thermally and chemically stable sp3 nanothread: poly [5] asterane},
  author={Saha, Biswajit and Pratik, Saied Md and Datta, Ayan},
  journal={Chemistry--A European Journal},
  volume={23},
  number={52},
  pages={12917--12923},
  year={2017},
  publisher={Wiley Online Library}
}

@article{wang2019two,
  title={Two-dimensional carbon-based auxetic materials for broad-spectrum metal-ion battery anodes},
  author={Wang, Shuaiwei and Si, Yubing and Yang, Baocheng and Ruckenstein, Eli and Chen, Houyang},
  journal={The journal of physical chemistry letters},
  volume={10},
  number={12},
  pages={3269--3275},
  year={2019},
  publisher={ACS Publications}
}

@article{baimova2017auxetic,
  title={AUXETIC BEHAVIOUR OF CARBON NANOSTRUCTURES.},
  author={Baimova, JA and Rysaeva, L Kh and Dmitriev, SV and Lisovenko, DS and Gorodtsov, VA and Indeitsev, DA},
  journal={Materials Physics \& Mechanics},
  volume={33},
  number={1},
  year={2017}
}

@article{gao2021new,
  title={New concept of carbon fiber reinforced composite 3D auxetic lattice structures based on stretching-dominated cells},
  author={Gao, Ying and Zhou, Zhengong and Hu, Hong and Xiong, Jian},
  journal={Mechanics of Materials},
  volume={152},
  pages={103661},
  year={2021},
  publisher={Elsevier}
}

@article{huang2016negative,
  title={Negative Poisson's ratio in modern functional materials},
  author={Huang, Chuanwei and Chen, Lang},
  journal={Advanced Materials},
  volume={28},
  number={37},
  pages={8079--8096},
  year={2016},
  publisher={Wiley Online Library}
}

@article{dmitriev2001poisson,
  title={Poisson ratio beyond the limits of the elasticity theory},
  author={Dmitriev, Sergey V and Shigenari, Takeshi and Abe, Kohji},
  journal={Journal of the Physical Society of Japan},
  volume={70},
  number={5},
  pages={1431--1432},
  year={2001},
  publisher={The Physical Society of Japan}
}

@article{lethbridge2010elastic,
  title={Elastic anisotropy and extreme Poisson’s ratios in single crystals},
  author={Lethbridge, Zoe AD and Walton, Richard I and Marmier, Arnaud SH and Smith, Christopher W and Evans, Kenneth E},
  journal={Acta Materialia},
  volume={58},
  number={19},
  pages={6444--6451},
  year={2010},
  publisher={Elsevier}
}

@article{grima2005auxetic,
  title={Auxetic behaviour from rotating rigid units},
  author={Grima, Joseph N and Alderson, Andrew and Evans, KE},
  journal={Physica status solidi (b)},
  volume={242},
  number={3},
  pages={561--575},
  year={2005},
  publisher={Wiley Online Library}
}

@article{greaves2011poisson,
  title={Poisson's ratio and modern materials},
  author={Greaves, George Neville and Greer, A Lindsay and Lakes, Roderic S and Rouxel, Tanguy},
  journal={Nature materials},
  volume={10},
  number={11},
  pages={823--837},
  year={2011},
  publisher={Nature Publishing Group}
}

@article{Grima.pssb.200460376,
author = {Grima, J. N. and Alderson, A. and Evans, K. E.},
title = {Auxetic behaviour from rotating rigid units},
journal = {physica status solidi (b)},
volume = {242},
number = {3},
pages = {561-575},
doi = {https://doi.org/10.1002/pssb.200460376},
year = {2005}
}

@article{doi:10.1073/pnas.1717442115,
author = {Daniel R. Reid  and Nidhi Pashine  and Justin M. Wozniak  and Heinrich M. Jaeger  and Andrea J. Liu  and Sidney R. Nagel  and Juan J. de Pablo },
title = {Auxetic metamaterials from disordered networks},
journal = {Proceedings of the National Academy of Sciences},
volume = {115},
number = {7},
pages = {E1384-E1390},
year = {2018},
doi = {10.1073/pnas.1717442115}
}

\end{document}